\documentclass[12pt]{article}
\usepackage{amssymb,amsfonts}
\usepackage{epsf,epsfig}
\begin{document}
\baselineskip 18pt

\title{Incoherent bound states in an infinite $XXZ$ chain at $\Delta=-1/2$}
\author{P.~N.~Bibikov\\ \it Sankt-Petersburg State University}

\maketitle

\vskip5mm

\begin{abstract}
For an infinite $XXZ$ chain with $\Delta=-1/2$ we have obtained a
family of translationary invariant three-magnon states which do
not satisfy the string conjecture. All of them have the same
energy.
\end{abstract}

\section{Introduction}

We shall study an infinite $XXZ$ spin chain \cite{1} related to
the Hamiltonian
\begin{equation}
H=\sum_{n=-\infty}^{\infty}H_{n,n+1},
\end{equation}
where
\begin{equation}
H_{n,n+1}={\bf S}^x_n{\bf S}^x_{n+1}+{\bf S}^y_n{\bf
S}^y_{n+1}+\Delta\Big({\bf S}^z_n{\bf S}^z_{n+1}-\frac{1}{4}\Big).
\end{equation}
The corresponding Hilbert space is an infinite tensor product of
${\mathbb C}^2$ spaces associated with the lattice sites. In every
such space we shall use the following basis
\begin{equation}
{\bf S}^z_n|\pm\rangle_n=\pm\frac{1}{2}|\pm\rangle_n.
\end{equation}
Here and in (2) ${\bf S}_n$ denotes a triple of $S=1/2$ spin
operators associated with $n$-th site. $\Delta$ is a real
parameter. The following transformation
\begin{equation}
H\rightarrow-UH U^{-1},
\end{equation}
where
\begin{equation}
U=\prod_n{\bf\sigma}_{2n}^z,
\end{equation}
(${\bf\sigma}^j_n=2{\bf S}_n^j$ for $j=x,y,z$ are the Pauli
matrices) is equivalent to the substitution
$\Delta\rightarrow-\Delta$. The result of our paper corresponds to
the special case $\Delta=-1/2$.

Traditionally the model (1) is treated on a finite chain related
to the Hilbert space $({\mathbb C}^2)^{\otimes N}$ ($N$ is the
number of sites). Usually there supposed periodic boundary
conditions \cite{2}-\cite{4}
\begin{equation}
H^{(period)}= \sum_{n=1}^NH_{n,n+1},\quad N+1\equiv1,
\end{equation}
(see however \cite{5} where the chain with open boundaries was
studied).

Since both the Hamiltonians (1) and (6) commute with ${\bf S}^z$
the $z$ component of the total spin
\begin{equation}
{\bf S}=\sum_n{\bf S}_n
\end{equation}
their spectrums split on subsectors corresponding to different
values of ${\bf S}^z$. Bethe Ansatze is used as an effective
method for treating the Hamiltonian (1) \cite{1} or (6)
\cite{2}-\cite{4}  separately in all subsectors. Within this
approach first of all is considered the highest ${\bf S}^z$ state
\begin{equation}
|\Omega\rangle=\prod_n|+\rangle_n,
\end{equation}
which is an eigenvector of both  (1) and (6). The next sector is
generated by quasiparticles (magnons). Since both the Hamiltonians
(1) and (6) commute with lattice translations one can readily
obtain an explicit form of the one-magnon state with quasi
momentum $k$
\begin{equation}
|1,k\rangle=\sum_n{\rm e}^{ikn}
\Big(\prod_{m=n_{min}}^{n-1}\otimes|+\rangle_m\Big)\otimes|-\rangle_n\otimes\Big(\prod_{m=n+1}^{n_{max}}\otimes|+\rangle_m
\Big),
\end{equation}
where for the infinite chain $n_{min}=-\infty$, $n_{max}=\infty$
while for the finite one $n_{min}=1$, $n_{max}=N$. The
corresponding dispersion
\begin{equation}
E_{magn}(k)=\cos{k}-\Delta,
\end{equation}
readily follows from the local action formulas
\begin{eqnarray}
H_{n,n+1}\dots|\mp\rangle_n|\pm\rangle_{n+1}\dots&=&
-\frac{\Delta}{2}\dots|\mp\rangle_n|\pm\rangle_{n+1}\dots+
\frac{1}{2}\dots|\pm\rangle_n|\mp\rangle_{n+1}\dots,\nonumber\\
H_{n,n+1}\dots|\pm\rangle_n|\pm\rangle_{n+1}\dots&=&0,
\end{eqnarray}
which are consequences of (2).

The exponent ${\rm e}^{ikn}$ in (9) is a one-magnon wave function.
Within Bethe Ansatze wave functions of all eigenstates are
represented as sums of Bethe exponents. For example for a general
two-magnon state
\begin{equation}
|2,k_1,k_2\rangle=\sum_{m<n}\psi(m,n;k_1,k_2)\dots|-\rangle_m\dots|-\rangle_n\dots,
\end{equation}
(where $\dots$ denote a product of suitable $|+\rangle_l$, $l\neq
m,n$ similar to the products in (9)) the wave function should be a
superposition of two Bethe exponents
\begin{equation}
\psi(m,n;k_1,k_2)=C_{12}(k_1,k_2){\rm
e}^{i(k_1m+k_2n)}+C_{21}(k_1,k_2){\rm e}^{i(k_2m+k_1n)}.
\end{equation}
The parameters $k_1$ and $k_2$ have sense of magnon quasi
momentums. They must be either real or complex. The former case
correspond to a scattering state while the later to a bound one.
Correspondingly a $n$-magnon wave function is a linear combination
of $n!$ Bethe exponents depending on $n$ different parameters
$k_1,\dots,k_n$. The corresponding total quasi momentum
$k=\sum_{j=1}^nk_j$ must be real. The dispersion is
\begin{equation}
E(k_1,\dots,k_n)=\sum_{j=1}^nE_{magn}(k_j).
\end{equation}

Both for (1) and (6) the parameters $k_j$ can not be arbitrary.
However the corresponding restrictions on them are different. In
the infinite case one should postulate that the wave function is
bounded. This requirement results in some inequalities on
imaginary parts of $k_j$ (see for example Eq. (18) below). In
other respects the parameters $k_j$ may be arbitrary. From the
opposite side in the finite case the parameters $k_j$ are
solutions of a transcendental system of Bethe equations. That is
why the finite problem is much more complicated than the infinite
one where the Bethe equations are not essential at all.

Of course physically relevant results usually belongs to infinite
chains. However within some approaches they may be obtained only
after suitable finite chain calculations before passing to the
$N\rightarrow\infty$ limit \cite{2},\cite{4}. Of course the latter
must be defined correctly. First of all one has to control
disappearance of all "bad" exponents resulting unbounded wave
functions. But there is another very important statement relevant
to $N\rightarrow\infty$ behavior of quasi momentums. Namely this
is the so called string conjecture which asserts that at
$N\rightarrow\infty$ all quasi momentums considered as solutions
of the Bethe equations group into special complexes "strings".
Within each of them all $k_j$ have similar real parts while their
imaginary parts form equidistant lattices symmetric with respect
to the real axis. For example for a two-magnon bound state related
to the wave function (13) there must be $k_1=u-iv$, $k_2=u+iv$
($v>0$). Since in (13) $m<n$ the second term is "bad" (unbounded).
So there should be $C_{21}(k_1,k_2)=0$. This condition produce a
relation between $u$ and $v$.

All magnons within the same complex have a {\it similar spatial
dependence of phase}. That is why a complex may be considered as a
{\it coherent} bound state of the corresponding magnons

Usually it is assumed that for the $XXZ$ model the string
conjecture is right. Within this assumption thermodynamics of the
infinite $XXZ$ chain was studied in \cite{6} for $|\Delta|\geq1$
and in \cite{7} for $|\Delta|<1$. However in the present paper we
show that the string conjecture fails in the special point
$\Delta=-1/2$. Namely we shall present a family of three-magnon
infinite-chain {\it incoherent} bound states with {\it total zero
quasi momentum}.

The $\Delta=-1/2$ $XXZ$ chain is now intensively studied in
various aspects (see the recent articles \cite{8}-\cite{10} and
references therein). We believe that our result shed an additional
light on this model.

\section{Three-magnon incoherent bound states}

First of all let us utilize the translation invariance and
represent a three magnon state with total quasimomentum $k$ in the
following general form
\begin{equation}
|3,k\rangle=\sum_{m<n<p} {\rm e}^{ik(m+n+p)/3}a(k,n-m,p-n)
\dots|-\rangle_m\dots|-\rangle_n\dots|-\rangle_p\dots.
\end{equation}
Reduced (to the center mass frame) wave function $a(k,m,n)$ has a
physical sense only at $m,n>0$ but may be continued to $m=0,\,n>0$
and $m>0,\,n=0$ according to Bethe conditions
\begin{eqnarray}
2\Delta a(k,1,n)&=&{\rm e}^{ik/3}a(k,0,n)+{\rm e}^{-ik/3}a(k,0,n+1),\nonumber\\
2\Delta a(k,m,1)&=&{\rm e}^{-ik/3}a(k,m,0)+{\rm
e}^{ik/3}a(k,m+1,0).
\end{eqnarray}
Under (16) the ${\rm Schr\ddot odinger}$ equation in the whole
region $m,n>0$ takes the following form
\begin{eqnarray}
&\displaystyle-3\Delta a(k,m,n)+\frac{1}{2}\Big[{\rm
e}^{-ik/3}a(k,m+1,n)+{\rm
e}^{ik/3}a(k,m-1,n)&\nonumber\\
&+{\rm e}^{-ik/3}a(k,m-1,n+1)+{\rm e}^{ik/3}a(k,m+1,n-1)+{\rm
e}^{-ik/3}a(k,m,n-1)&\nonumber\\
&+{\rm e}^{ik/3}a(k,m,n+1)\Big]=Ea(k,m,n).&
\end{eqnarray}

The following trial bounded wave function
\begin{equation}
a(m,n)={\rm e}^{(iu_1-v_1)m+(iu_2-v_2)n},
\end{equation}
satisfy (17) for
\begin{eqnarray}
E(k,u_1,u_2,v_1,v_2)&=&\cosh{v_1}\cos{(k/3-u_1)}+\cosh{v_2}\cos{(k/3+u_2)}\nonumber\\
&+&\cosh{(v_1-v_2)}\cos{(k/3+u_1-u_2)}-3\Delta.
\end{eqnarray}
Normalization condition
\begin{equation}
\sum|a(m,n)|^2<\infty,
\end{equation}
results in
\begin{equation}
v_{1,2}>0.
\end{equation}

Form the other side the system (16) gives
\begin{eqnarray}
x_1&=&{\rm e}^{i(k/3+u_1)-v_1}+{\rm e}^{-ik/3}-2\Delta_1{\rm
e}^{iu_2-v_2}=0,\nonumber\\
x_2&=&{\rm e}^{ik/3}+{\rm e}^{i(u_2-k/3)-v_2}-2\Delta_1{\rm
e}^{iu_1-v_1}=0.
\end{eqnarray}

Treating $x_1-\bar x_2$ one may readily obtain
\begin{equation}
2{\Delta}F=-\bar F{\rm e}^{ik/3},
\end{equation}
where $F={\rm e}^{iu_2-v_2}-{\rm e}^{-iu_1-v_1}$.

At $4\Delta^2\neq1$ Eq. (23) gives $F=0$ or equivalently
$u_1=-u_2$ and $v_1=v_2$. In this case the string conjecture is
satisfied. However in two special points $\Delta=\pm1/2$ connected
by the symmetry (4) there should be additional solutions. Taking
$\Delta=-1/2$ and treating $x_1-{\rm e}^{ik/3}x_2$ one gets
\begin{equation}
k=0.
\end{equation}
Now the system (22) results in
\begin{equation}
{\rm e}^{-v_1}\cos{u_1}+{\rm e}^{-v_2}\cos{u_2}=-1,\quad{\rm
e}^{-v_1}\sin{u_1}=-{\rm e}^{-v_2}\sin{u_2},
\end{equation}
or
\begin{equation}
{\rm e}^{v_1}=\frac{\sin{(u_1-u_2)}}{\sin{u_2}},\quad{\rm
e}^{v_2}=\frac{\sin{(u_2-u_1)}}{\sin{u_1}}.
\end{equation}
According to (26) $\sin{u_1}$ and $\sin{u_2}$ have opposite signs.
Without loss of generality one may put
\begin{equation}
0<u_1<\pi,\quad -\pi<u_2<0.
\end{equation}
Under this assumption both $\cos{u_{1,2}/2}>0$ and the system (21)
is reducible to
\begin{equation}
\sin{\Big(u_1-\frac{u_2}{2}\Big)}<0,\quad
\sin{\Big(\frac{u_1}{2}-u_2\Big)}<0,
\end{equation}
or equivalently
\begin{equation}
\pi<u_1-\frac{u_2}{2}<2\pi,\quad\pi<\frac{u_1}{2}-u_2<2\pi.
\end{equation}

It may be readily proved from (19) and (26) that all these states
have zero energy. According to (15) and (18) they describe magnon
triples with corresponding quasi momentums
\begin{equation}
k_1=-u_1-iv_1,\quad k_2=u_1-u_2+i(v_1-v_2),\quad k_3=u_2+iv_2.
\end{equation}
The string conjecture obviously is failed.

The author thanks P. P. Kulish for careful reading of the
manuscript.

\end{document}